\documentclass[reprint,amsmath,amssymb,aps,pra,showpacs]{revtex4-1}

\usepackage{graphicx}
\usepackage{dcolumn}
\usepackage{bm}
\usepackage[utf8]{inputenc}
\usepackage{amsfonts}
\usepackage{tabularx}
\usepackage[normalem]{ulem}

\usepackage{braket} 
\usepackage{nicefrac} 
\usepackage{IEEEtrantools} 

\newcount\colveccount
\newcommand*\colvec[1]{
        \global\colveccount#1
        \begin{pmatrix}
        \colvecnext
}
\def\colvecnext#1{
        #1
        \global\advance\colveccount-1
        \ifnum\colveccount>0
                \\
                \expandafter\colvecnext
        \else
                \end{pmatrix}
        \fi
}

\begin{document}
\title{Higher-order corrections for the dynamic hyperfine structure of muonic atoms}

\author{Niklas Michel}\email{nmichel@mpi-hd.mpg.de}
\author{Natalia S. Oreshkina}

\affiliation{Max~Planck~Institute for Nuclear Physics, Saupfercheckweg 1,
69117 Heidelberg, Germany}

\date{\today}


\begin{abstract}
A method for precise calculation of the energy corrections due to second order electric quadrupole interactions, as well as mixed electric {quadrupole-vacuum} polarization  in the framework of the dynamic hyperfine structure in heavy muonic atoms is presented. For this, a multipole expansion of the Uehling potential is performed. The approach is applicable for an arbitrary nuclear electric charge distribution. By performing these calculations for muonic Rhenium and Uranium using a deformed Fermi distribution, it is shown that both corrections contribute on a level presumably visible in upcoming experiments.
\end{abstract}

\maketitle

\section{Introduction}
A muon is an elementary particle which is similar to the electron in many aspects, in particular it has the same electric charge. Bound states between a muon and an atomic nucleus are commonly referred to as muonic atoms. The muon, however, is about two hundred times heavier than the electron~\cite{codata2016} and the Bohr radius of muonic atoms is correspondingly downsized by the same factor. If compared to electronic bound states, the wave functions of muonic bound states therefore have a much bigger overlap with the nucleus~\cite{Patoary2018}, and it has been recognized early that muonic transition energies enable the extraction of information about the nuclear charge distribution~\cite{wheeler1953}. In a typical experimental setting, atomic electrons are also present and electron-muon interaction has to be taken into account in principle. However, it was shown that the screening effect due to atomic electrons is small for low-lying muonic states~\cite{michel2017,vogel1973,fricke1969} and therefore muonic atoms can essentially be considered as a hydrogen-like system. In contrast to electronic atoms, where the magnetic dipole splitting dominates over the electric quadrupole splitting, the muonic magnetic dipole splitting is suppressed because the magnetic moment of the muon is smaller than the electron's one by a factor of the electron-muon mass ratio. Therefore, the hyperfine splitting in muonic atoms is mainly due to electric quadrupole interaction, providing an ideal environment for testing electric interactions. Experiments on muonic atoms have already provided nuclear parameters like RMS radii~\cite{fricke1995} and quadrupole moments~\cite{stone2016,tanaka1983}, as well as knowledge about the distribution of electric charge inside the nuclei, e.g.~\cite{tanaka1984_2,tanaka1984}, for a wide range of charge numbers. Also, experiments on muonic and electronic hydrogen resulted in a different value of the proton radius~\cite{Pohl2010}. For the extraction of nuclear parameters a thorough understanding of the muon spectrum for a given nuclear model is indispensable~\cite{BorieRinker1982}. This includes, amongst others, the influence of a spatially extended distribution of electric charge and quadrupole moment inside the nucleus. Especially for heavy nuclei, the quadrupole interaction between muon and nucleus beyond first order can be important~\cite{tanaka1984,hitlin1970,wu1969,Devons1995}, which is called the dynamic hyperfine structure in muonic atoms. As a consequence, transition energies have to be calculated by diagonalizing the quadrupole interaction in a small modelspace. A non-relativistic estimation of the residual second order electric quadrupole interaction with states outside of this modelspace has been presented earlier in Ref.~\cite{chen1970}.

The contributions from bound state quantum electrodynamics also influence muonic atoms significantly. In particular, the vacuum polarization (VP) by virtual electron-positron pairs modifies the electric interaction between muon and nucleus at short distances. Due to the small electron-muon mass ratio, the electronic loop is far less suppressed compared to electronic bound states and leads to sizeable corrections of the energy levels in muonic atoms. VP influences multipole interactions of all orders. In the past, however, the corresponding correction to the quadrupole interaction between muon and nucleus has at most been considered with a power-series expansion~\cite{Fricke1969vp,zehnder1975} or for specific forms of the nuclear charge distribution~\cite{pearson1963}, which does not enable precision calculations for heavy muonic atoms.

With upcoming experiments on high-Z muonic atoms~\cite{kirch2016} and anticipating increasing experimental precision, an accurate treatment of the quadrupole interactions including leading order VP and second order terms is desirable. To tackle this shortcoming, we derive the VP correction in Uehling approximation for multipole interactions of any order and for arbitrary charge distributions. Thereby, a numerical treatment of quadrupole interactions in heavy muonic atoms up to second order including nuclear finite size effects and VP is presented, using relativistic muon states. It is shown that it can lead to energy shifts potentially important for the extraction of nuclear parameters from future experiments.

Muonic relativistic units with ${\hbar}{=}{c}{=}{m_\mu}{=}{1}$ are used throughout this work, where $m_\mu$ is the muon's mass, along with the Heavyside charge unit with ${\alpha}{=}{e^2}{/}{4\pi}$, where $\alpha$ is the fine structure constant and the elementary charge $e$ is negative.\\

\section{Rotational Nuclear Model for \\Muonic Atoms}
The typical energy scale for nuclear transitions is a few orders
of magnitude larger than for an electronic transitions
in atoms. Therefore, the electrons essentially interact only
with the nuclear ground state. In muonic atoms, however, the energy scale of the hyperfine structure can be of the same order as low lying nuclear rotational states.
Therefore, in muonic atoms a high degree of
mixing between muonic and nuclear states is present and the excitation energies and quadrupole moments of several nuclear states are needed for the description of the level structure. Nuclear parameters can be taken from experimental data or, if not available, obtained by theoretical models. The rotational nuclear model for muonic atoms has been very successful in describing the level structure of heavy muonic atoms with large electric quadrupole interactions between muon and atomic nucleus~\cite{tanaka1984,hitlin1970,wu1969,Devons1995}. The nucleus is modeled as a rigid rotor, where a fixed distribution of electric charge, given in the nuclear frame, rotates in the laboratory frame. The muon is described as a Dirac particle coupled to such a nucleus. Thereby, the Hamiltonian of the coupled muon-nucleus system reads
\begin{equation}
H = H_{\text{N}} + H_\mu + V_{\text{el}} + V_{\text{uehl}},
\label{eq:htot}
\end{equation}
where $H_{\text{N}}$ is the nuclear Hamiltonian, and ${H_\mu}{=}{\vec{\alpha} \cdot \vec{p} + \beta m_\mu}$ is the free Dirac Hamiltonian for the muon with momentum $\vec{p}$, and $\vec{\alpha}$ and $\beta$ are the four Dirac matrices. 
The interaction between muon and nucleus is described by including the electric interaction potential $V_{\text{el}}$ and additionally the VP correction in Uehling approximation $V_{\text{uehl}}$ in the Hamiltonian.

The degrees of freedom of the rigid rotor model are the three Euler angles $\phi$, $\theta$, and $\psi$ describing the position of the nuclear body fixed frame in the laboratory frame. The corresponding wave functions for the rotational part are proportional to the Wigner functions $D^{I^{\,*}}_{MK}(\phi,\theta,\psi)$ or $\left|IMK\right>$ in braket notation~\cite{ring_schuck,brown_carrington}. $I$ is the total nuclear angular momentum, $M$ and $K$ are the projections on the $z$ axis of the laboratory and nuclear body-fixed system, respectively. Throughout this work, the conventions and notation of the angular momentum algebra, like Wigner $D$ functions, Clebsch-Gordan coefficients, and $6j$~symbols, correspond to~\cite{varshalovich1988}. The energies of the nuclear rotational states
\begin{equation}
H_{\text{N}} \left|IMK\right> = E_I \left|IMK\right>
\label{eq:enucl}
\end{equation}
are typically taken from literature~\cite{ENSDF}. The electric interaction in terms of the nuclear charge distribution $\rho(\vec{r})$ reads
\begin{equation}
V_{\text{el}}(\vec{r}_\mu^{\,\prime}) = -Z \alpha \int \text{d}^3 r_N^{\prime}\, \frac{\rho(\vec{r}_N^{\,\prime})}{|\vec{r}_\mu^{\,\prime}-\vec{r}_N^{\,\prime}|},
\label{eq:pot}
\end{equation}
where primed coordinates belong to the nuclear system and unprimed to the laboratory system, ${\vec{r}_{\mu}}{=}{(r_\mu,\vartheta_\mu,\varphi_\mu)}$ are the muonic coordinates. A multipole expansion of Eq.~\eqref{eq:pot} in the laboratory frame has been demonstrated eg. in Ref.~\cite{kozhedub2008}, so we refer to Appendix \ref{sec:multipole} for derivations and explicit expressions, and denote the result as
\begin{equation}
V_{\text{el}}(\vec{r}_\mu,\phi,\theta) = \sum_{l=0}^\infty V^{(l)}_{\text{el}}(\vec{r}_\mu,\phi,\theta).
\label{eq:multipoles}
\end{equation}
For a nuclear charge distribution with axial and reflection symmetry, the terms with odd~$l$ vanish, thus the first two non-vanishing terms are the monopole (${l}{=}{0}$) and quadrupole (${l}{=}{2}$) term.

To first order  in $\alpha$ and $Z\alpha$, VP due to a virtual electron-positron pair can be expressed explicitly in position space for an arbitrary nuclear charge distribution and is referred to as Uehling potential, which reads in the chosen system of units as~\cite{Fullerton1976}
\begin{equation}
{V_{\text{uehl}}(\vec{r}_\mu^{\,\prime})}{=}{-Z\alpha\frac{2 \alpha}{3\pi} \int \text{d}^3r_N^{\prime} \frac{\rho(\vec{r}_N^{\,\prime})}{|\vec{r}_\mu^{\,\prime} - \vec{r}_N^{\,\prime}\,|} K_1(2m_e{|\vec{r}_\mu^{\,\prime} - \vec{r}_N^{\,\prime}\,|}),}
\label{eq:Vvp}
\end{equation}
where $m_e$ is the electron mass and $K_1(x)$ belongs to the family of functions
\begin{equation}
K_n(x)=\int_1^\infty \text{d}t\,\text{e}^{-xt}\left(\frac{1}{t^3}+\frac{1}{2t^5}\right)\sqrt{t^2-1}t^n.
\label{eq:defKn}
\end{equation}
In order to obtain the VP corrections to the quadrupole interaction, a multipole expansion of Eq.~(\ref{eq:Vvp}) has to be performed in a similar way, whereas the dependence on $|\vec{r}_\mu^{\,\prime}-\vec{r}^{\,\prime}|$ now also is present in the argument of $K_1(x)$. The result in the laboratory frame can be written similar to Eq.~\eqref{eq:multipoles} as
\begin{equation}
V_{\text{uehl}}(\vec{r}_\mu,\phi,\theta) = \sum_{l=0}^\infty V^{(l)}_{\text{uehl}}(\vec{r}_\mu,\phi,\theta),
\label{eq:uehlmultipoles}
\end{equation}
where the ${l}{=}{0}$-term is the well-known Uehling potential for a spherically symmetric charge distribution, given in Eq.~\eqref{eq:sph_uehl}, and the ${l}{=}{2}$-term is the corresponding correction of the quadrupole interaction. The derivation and expressions for $V^{(l)}_{\text{uehl}}(\vec{r}_\mu,\phi,\theta)$ in Eq.~\eqref{eq:uehlmultipoles} can be found in Appendix \ref{sec:multipole}.
%
%
\begin{figure}[b]
\includegraphics[width=0.22\textwidth]{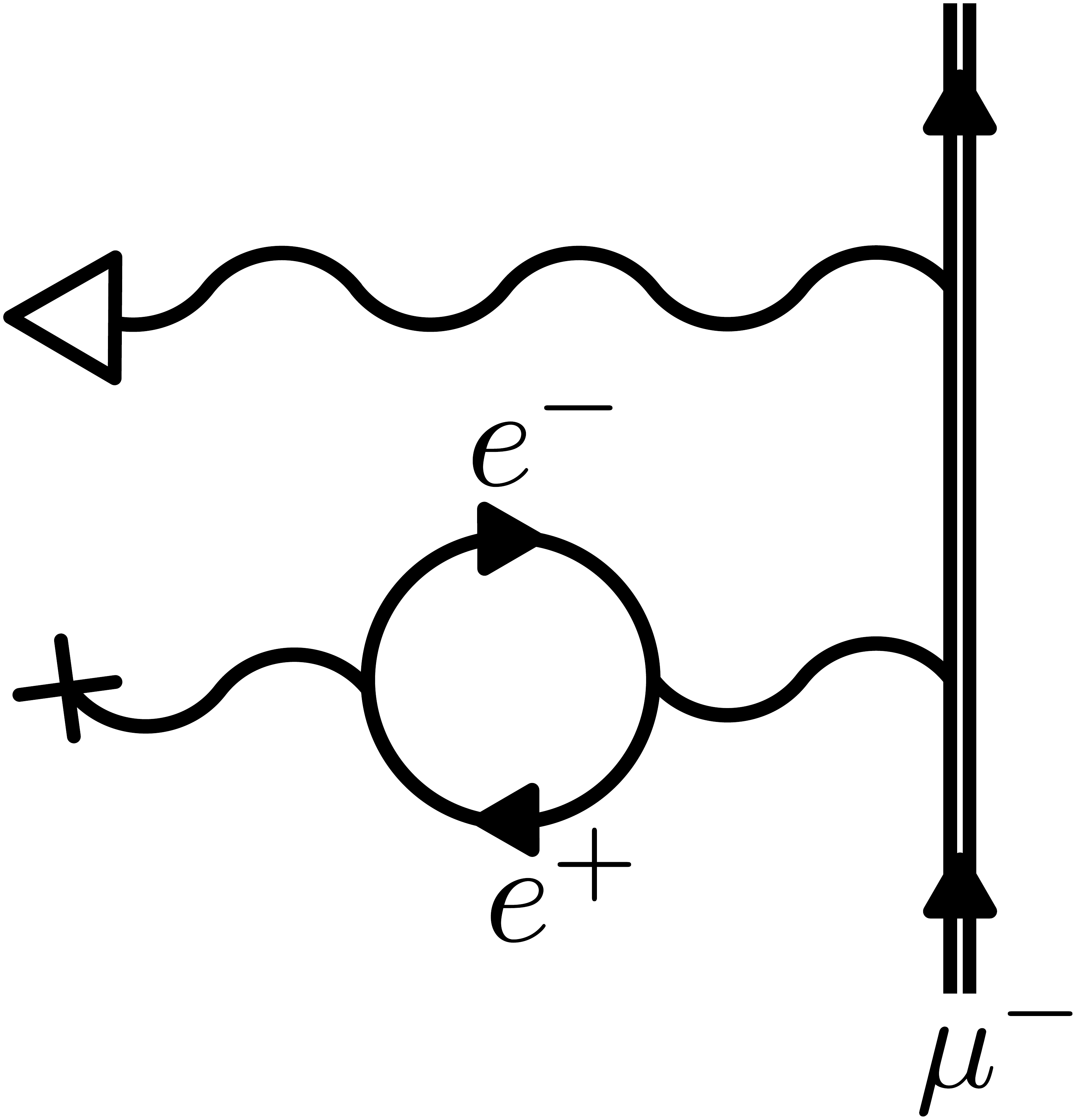}\hfill
\includegraphics[width=0.22\textwidth]{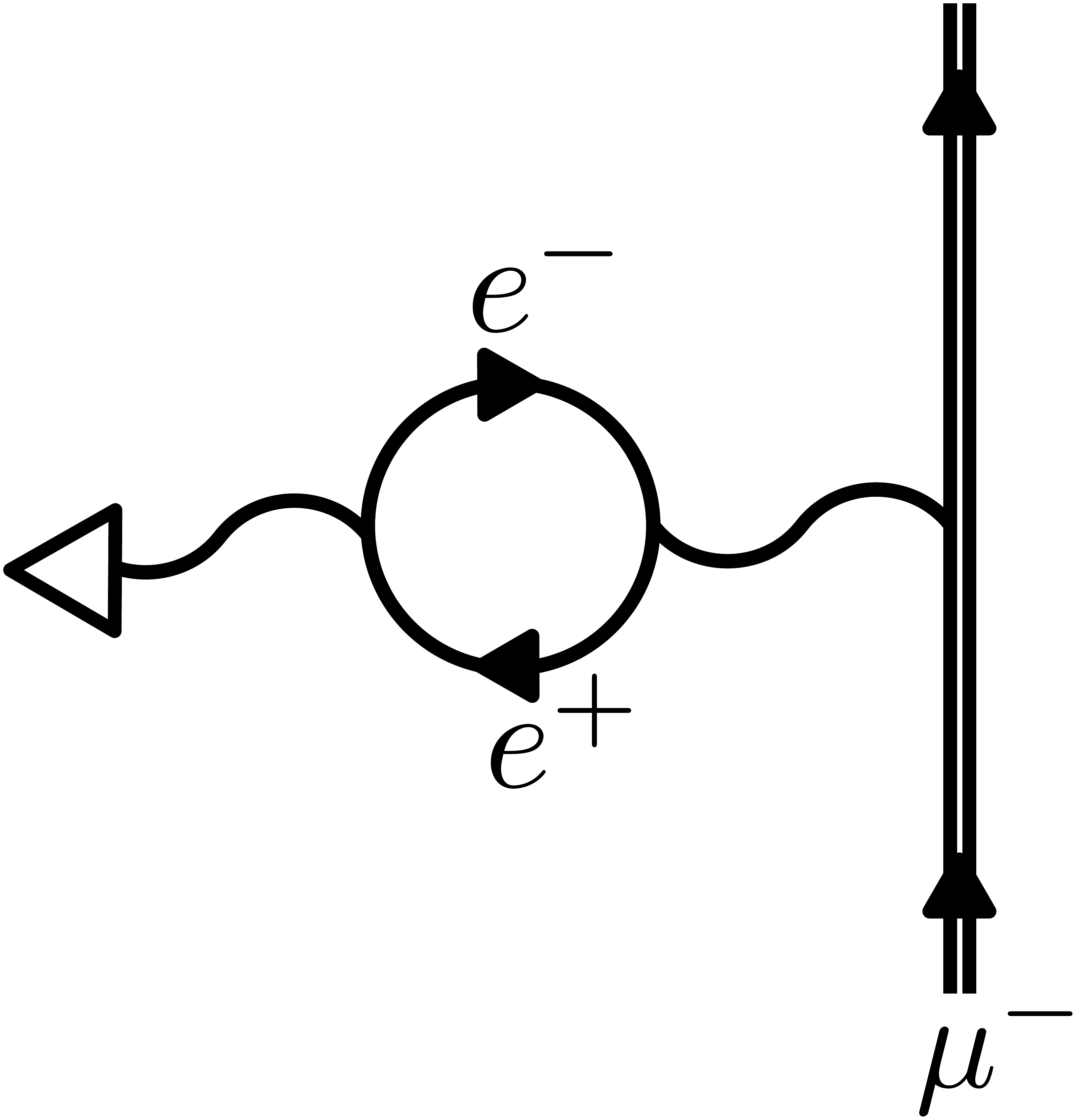}\\
$\qquad\qquad\qquad$(a)\hfill(b)$\qquad\qquad\qquad$
\caption{
Feynman diagrams for the leading order contributions of the vacuum polarization to the quadrupole interaction in muonic atoms. An external double line stands for the bound-muon wave function. A single internal line for the free electron propagator, and a wave line for the photon propagator. A cross represents the interaction with the monopole potential, and a triangle the interaction with the quadrupole potential. Contribution (a) is calculated by including the spherically symmetric contribution of the Uehling potential in the Dirac equation~\eqref{eq:h0mu}; Contribution (b) by including the quadrupole contribution of the Uehling potential in the matrix elements from Eq.~\eqref{eq:quadel_uehl} and \eqref{eq:second}.
}
\label{fig:quehl}
\end{figure}

The monopole terms $V^{(0)}_{\text{el}}(r_\mu)$ and $V^{(0)}_{\text{uehl}}(r_\mu)$ only depend on the muonic radial coordinate and thus were included in this work in the unperturbed muonic states as
\begin{equation}
\left( H_\mu + V^{(0)}_{\text{el}}+V^{(0)}_{\text{uehl}}\right) \left|n\kappa m\right> = E_{n\kappa}\left|n\kappa m \right>,
\label{eq:h0mu}
\end{equation}
where $\left|n\kappa m \right>$ are the solutions of the spherical Dirac equation in terms of the well-known spherical spinors and radial functions $G_{n\kappa}(r_\mu)$ and $F_{n\kappa}(r_\mu)$ with energy $E_{n \kappa}$~\cite{greiner2000}. $n$, $\kappa$, $m$ are the principal quantum number, the relativistic angular quantum number and the $z$ component of the total angular momentum of the muon, respectively. In this way, the monopole Uehling potential and all iterations thereof are included in the muonic wave functions~\cite{michel2017}.
\section{Electric quadrupole interaction}
\subsection{Dynamic Hyperfine structure}
The hyperfine splitting due to electric quadrupole interactions is considered in the coupled basis, where a state with total angular momentum $F$ and its projection on the $z$  axis $M$ is formed for a given muonic and nuclear state as
\begin{equation}
\left| FM\,n\kappa\,IK \right> = \sum_{m,M^\prime} C^{FM}_{jm\,IM^\prime} \left|n\kappa m\right>\otimes\left|IM^\prime K\right>,
\end{equation}
with the Clebsch-Gordan coefficients $C^{jm}_{j_1m_1\,j_2m_2}$. The matrix elements of the electric quadrupole interaction are written as
\begin{equation}
\Delta E^{(2)}_{\text{el}} = \left< FMn_1\kappa_1I_1K\right|{V_{\text{el}}^{(2)}}\left|FMn_2\kappa_2I_2K\right>,
\label{eq:quadel}
\end{equation}
where the diagonal elements correspond to the first order electric hyperfine splitting~\cite{michel2017}. The explicit expressions are given in Eqs.~\eqref{eq:matel} and~\eqref{eq:defmulti} in the Appendix.
\begin{table}[b]
\caption{\label{tab:params}%
Nuclear parameters used in the numerical calculations. $I_0$ is the nuclear ground state angular momentum. RMS and $Q_\text{spec}$ are the nuclear RMS radius and spectroscopic quadrupole moment of the nuclear ground state from~\cite{Angeli2013,Stone2005}, respectively. $c$, $a$, $\beta$ are the parameters of the deformed Fermi distribution derived from RMS and $Q_\text{spec}$, see Sec. \ref{sec:num} for details. $E_{I}$ are the excitation energies of the nuclear rotational states with angular momentum $I$ in Eq.~\eqref{eq:enucl}, the values are taken from~\cite{ENSDF}.}
\begin{ruledtabular}
\begin{tabular}{rll}
& $^{185}_{\phantom{1}75}$Re & $^{235}_{\phantom{1}92}$U\\ \hline \\[-10pt]
$I_0$ \hfill$\phantom{.}$ & $5/2$ & $7/2$ \\
RMS \hfill[fm] & $5.3596(172)$ & $5.8337(41)$ \\
$Q_\text{spec}$ \hfill[b] & $2.21(4)\phantom{111}$ & $4.936(6)\phantom{1}$ \\
$c$ \hfill[fm] & $6.3517$ & $6.9562$ \\
$a$ \hfill[fm] & 0.5234 & 0.5234 \\
$\beta$ \hfill$\phantom{.}$ & 0.2322 & 0.2711 \\[7pt]
$E_{I_0 + 1}$ \hfill[keV] & 125.3587(9) &  $\phantom{1}$46.108(8) \\
$E_{I_0 + 2}$ \hfill[keV] & 284.2(3) & 103.903(8) \\
$E_{I_0 + 3}$ \hfill[keV] & 475.7(4) & 171.464(13) \\
$E_{I_0 + 4}$ \hfill[keV] & 697.1(5) & 250.014(21) \\
$E_{I_0 + 5}$ \hfill[keV] & 949.7(5) & 339.976(24) \\
\end{tabular}
\end{ruledtabular}
\end{table}
For heavy, deformed nuclei, the fine structure as well as the first order quadrupole hyperfine structure of muonic energy levels is similar to the energies of the low-lying nuclear rotational states. Therefore, the total Hamiltonian has to be rediagonalized in finite subspaces of a muonic finestructure multiplet and the first few nuclear rotational states. Since multipole interactions are diagonal in $F$ and $M$, there is no mixing of states with different total angular momentum. Thus, if $d$ is the dimension of the subspace, new states
\begin{equation}
\left| FM k \right> := \sum_{i=1}^{d} a^{(k)}_i \left| FM\,n_i\kappa_i\,I_iK \right>
\label{eq:compstate}
\end{equation}
with $k \in {1,...,d}$ may be defined, in which the coefficients $a_i^{(k)}$ have to be chosen such that the matrix representation of the total Hamiltonian \eqref{eq:htot} in the subspace is diagonal. The corresponding eigenenergies are written as $E^{(F,k)}_{\text{quad}}$. This leads to a rich and complex level structure and to hyperfine structure also for nuclei with zero ground-state angular momentum and is know as the dynamic hyperfine structure~\cite{wu1969,Devons1995}.

\subsection{Higher order contributions}
In this work, the VP in Uehling approximation is included in the dynamic hyperfine structure, considering the finite nuclear size. For this, the matrix elements of the quadrupole Uehling interaction
\begin{equation}
\Delta E^{(2)}_{\text{uehl}} = \left< FMn_1\kappa_1I_1K\right|{V_{\text{uehl}}^{(2)}}\left|FMn_2\kappa_2I_2K\right>,
\label{eq:quadel_uehl}
\end{equation}
are added to the matrix elements from Eq.~\eqref{eq:quadel}. The explicit formulas for Eq.~\eqref{eq:quadel_uehl} are given in Eqs.~\eqref{eq:matel} and~\eqref{eq:defmultiuehl} in the Appendix.
By including the monopole Uehling potential in the unperturbed muon states in Eq.~\eqref{eq:h0mu} and the quadrupole Uehling potential in the matrix elements~\eqref{eq:quadel_uehl}, all contributions of the Uehling potential to the electric quadrupole interactions are included, as shown schematically in Fig.~\ref{fig:quehl}~(a) and (b), respectively.

After a subspace has been chosen and rediagonalization has been performed, the quadrupole interaction with states outside of the subspace leads to residual second order corrections to the energy levels~\cite{chen1970}. 
For the total second order correction a summation over the complete (discrete and continous) spectrum for both nuclear and muonic states has to be performed. For the complete nuclear spectrum, sophisticated models or numerous experimental data are required, causing the nuclear polarization corrections of the energy levels~\cite{nefiodov1996,chen1970,haga2002}. In this work, we calculate the second order corrections, where the nucleus stays in the rotational ground state, but the complete muonic spectrum is considered.
The second order energy shift is
\begin{equation}
\Delta E_{\text{2.ord.}}^{(F,k)}= \sum_i\frac{\left|\left< FMk\right|{V_{\text{el}}^{(2)}}{+}{V_{\text{uehl}}^{(2)}}\left|FMn_i\kappa_iI_iK \right>\right|^2}{E_{F,k}-E_i},
\label{eq:second}
\end{equation}
where the sum is to be taken over all states not considered in the rediagonalization, including continuum states of the muon, and the unperturbed energy of the state $i$ is $E_i=E_{n_i\kappa_i}+E_{I_i}$.
\section{Numerical Evaluation}
\label{sec:num}
Calculations have been performed for muonic Rhenium and Uranium, assuming a deformed Fermi nuclear charge distribution which reads as
\begin{equation}
\rho_{ca\beta}(\vec{r}\,)=\cfrac{N}{1+\exp (\frac{r-c(1+\beta \text{Y}_{20}(\vartheta))}{a})},
\label{eq:deffermi}
\end{equation}
where $c$ is the half-density radius, $a$ the skin thickness, $\beta$ the deformation parameter, and $N$ a normalization constant determined by the condition
\begin{equation}
\int \text{d}^3r\, \rho_{ca\beta}(\vec{r}\,)=1.
\end{equation}
Using the deformed Fermi distribution has proved to be very successful in the description of the level structure of heavy muonic atoms, e.g.~\cite{hitlin1970,tanaka1984,tanaka1984_2}.
Values for the parameters can be estimated by using a value ${a}{=}{2.3\,\text{fm}/(4\ln 3)}$, which has proved to be a sufficiently accurate value for most nuclei~\cite{Beier2000}. Then, $c$ and $\beta$ are chosen such that the quadrupole moment and RMS value of the distribution are in agreement with the literature values from~\cite{Angeli2013,Stone2005}. For the nuclear states involved in the dynamic hyperfine structure, also the excitation energies are needed from literature~\cite{ENSDF}. The used parameters are summarized in Table~\ref{tab:params}.
With these parameters, the electric and Uehling potentials, both monopole and quadrupole parts, can be calculated numerically. The muon wave functions are obtained by solving the Dirac equation~\eqref{eq:h0mu} with the dual-kinetic-balance method~\cite{Shabaev2004}. Thereby, a complete set of muonic bound and continuum states is obtained. An overview for the binding energies of muon states important for the dynamic hyperfine splitting are shown in Table~\ref{tab:monopole}.

\begin{table}[b]
\caption{\label{tab:monopole}%
Binding energies of the low-lying, unperturbed muonic states due to the spherically symmetric parts of the electric and Uehling potential obtained by solving Eq.~\eqref{eq:h0mu} for muonic Rhenium and Uranium. The first column shows the binding energies for a point-like Coulomb potential, the second and third column include the finite size corrections without and with Uehling potential, respectively. See Sec. \ref{sec:num} for details. All energies are in keV.
}
\begin{ruledtabular}
\begin{tabular}{ccrrr}
& state & \text{point like}& \text{finite size (fs)} &\text{fs+Uehling}\\ \hline \\[-7pt]
$^{185}$Re &1s\nicefrac{1}{2} & 17229.12 & 9333.46 & 9394.02 \\
&2s\nicefrac{1}{2} & 4398.85 & 3083.91 & 3100.44 \\
&2p\nicefrac{1}{2} & 4398.85 & 4032.61 & 4059.50 \\
&2p\nicefrac{3}{2} & 4033.07 & 3885.75 & 3910.50 \\
&3s\nicefrac{1}{2} & 1912.97 & 1498.01 & 1504.28 \\
&3p\nicefrac{1}{2} & 1912.97 & 1789.84 & 1798.66 \\
&3p\nicefrac{3}{2} & 1804.01 & 1751.38 & 1759.75 \\
&3d\nicefrac{3}{2} & 1804.01 & 1802.05 & 1810.30 \\
&3d\nicefrac{5}{2} & 1773.14 & 1772.36 & 1780.16 \\[7pt]
$^{235}$U&1s\nicefrac{1}{2} & 27351.29 & 12100.56 & 12175.51 \\
&2s\nicefrac{1}{2} & 7074.68 & 4308.67 & 4332.13 \\
&2p\nicefrac{1}{2} & 7074.68 & 5901.35 & 5941.39 \\
&2p\nicefrac{3}{2} & 6130.65 & 5674.78 & 5711.89 \\
&3s\nicefrac{1}{2} & 3033.18 & 2148.86 & 2158.31 \\
&3p\nicefrac{1}{2} & 3033.18 & 2645.58 & 2659.26 \\
&3p\nicefrac{3}{2} & 2751.54 & 2588.19 & 2601.27 \\
&3d\nicefrac{3}{2} & 2751.54 & 2739.69 & 2754.06 \\
&3d\nicefrac{5}{2} & 2679.66 & 2674.77 & 2688.10

\end{tabular}
\end{ruledtabular}
\end{table}
%
\begin{table*}
\caption{\label{tab:hfs}
Overview of energy corrections due to residual second order electric quadrupole splitting~$\Delta E_{\text{2.ord.}}$ and quadrupole-Uehling interaction~$\Delta E_{\text{quad-uehl}}$ for $^{185}$Re and $^{235}$U. $F$ is the total angular momentum of muon and nucleus, $I_N$ is the nuclear angular momentum and $\mu$-state is the muonic state in spectroscopic notation. For the muonic $2p$ and $3d$ states, these are mixed by the dynamic hyperfine structure, thus $I_N$ (main) and $\mu$-state (main) show the states with the largest contribution. $E_{\text{quad}}$ is the binding energy without quadrupole Uehling and residual second order quadrupole interaction, see Sec. \ref{sec:num} for details. The states are ordered descending in the total energy~$E_{\text{tot}}$. All energies are in keV.}
\begin{ruledtabular}
\begin{tabular}{lrccddd|d}
 &F&\multicolumn{1}{c}{$I_{N}\,\text{(main)}$}&$\mu\text{-state (main)}$&\multicolumn{1}{c}{$E_{\text{quad}}$}&\multicolumn{1}{c}{$\Delta E_{\text{2.ord.}}$}&\multicolumn{1}{c}{$\Delta E_{\text{quad-uehl}}$}&\multicolumn{1}{c}{$E_{\text{tot}}$}\\\hline\\[-7pt]
$^{185}\text{Re}$&  2 &   \nicefrac{5}{2} & 1s\nicefrac{1}{2} & 9394.02 &  3.21 &   0.00 & 9397.23 \\
&  6 &  \nicefrac{13}{2} & 1s\nicefrac{1}{2} & 8696.92 &  2.06 &   0.00 & 8698.98 \\
&  8 &  \nicefrac{15}{2} & 1s\nicefrac{1}{2} & 8444.32 &  1.76 &   0.00 & 8446.08 \\
&  2 &   \nicefrac{5}{2} & 2p\nicefrac{1}{2} & 4083.31 &  2.18 &  0.28 & 4085.77 \\
&  3 &   \nicefrac{5}{2} & 2p\nicefrac{1}{2} & 4077.79 &  2.07 &  0.23 & 4080.09 \\
&  3 &   \nicefrac{9}{2} & 2p\nicefrac{3}{2} & 3992.27 &  2.41 &  0.41 & 3995.09 \\
&  4 &   \nicefrac{7}{2} & 2p\nicefrac{1}{2} & 3957.33 &  2.10 &  0.26 & 3959.69 \\
&  3 &   \nicefrac{5}{2} & 2p\nicefrac{3}{2} & 3886.35 &  1.12 & -0.22 & 3887.25 \\
&  5 &   \nicefrac{7}{2} & 2p\nicefrac{3}{2} & 3814.27 &  2.08 &  0.28 & 3816.63 \\
&  4 &   \nicefrac{9}{2} & 2p\nicefrac{1}{2} & 3734.93 &  1.03 & -0.27 & 3735.69 \\
&  6 &   \nicefrac{9}{2} & 2p\nicefrac{3}{2} & 3650.57 &  1.95 &  0.25 & 3652.77 \\
&  5 &   \nicefrac{9}{2} & 2p\nicefrac{3}{2} & 3556.36 &  1.13 & -0.24 & 3557.25 \\
&  7 &  \nicefrac{11}{2} & 2p\nicefrac{3}{2} & 3458.14 &  1.85 &  0.23 & 3460.22 \\
&  6 &  \nicefrac{11}{2} & 2p\nicefrac{3}{2} & 3344.35 &  0.93 & -0.19 & 3345.09 \\
&  8 &  \nicefrac{13}{2} & 2p\nicefrac{3}{2} & 3111.03 &  0.68 &  0.02 & 3111.73 \\
&  7 &  \nicefrac{15}{2} & 2p\nicefrac{3}{2} & 2941.66 &  0.82 & -0.15 & 2942.33 \\
&  8 &  \nicefrac{15}{2} & 2p\nicefrac{3}{2} & 2938.52 &  0.67 & -0.16 & 2939.03 \\
&  3 &   \nicefrac{5}{2} & 3d\nicefrac{3}{2} & 1815.47 &  0.07 &  0.03 & 1815.57 \\
&  1 &   \nicefrac{5}{2} & 3d\nicefrac{3}{2} & 1804.28 &  0.11 & -0.03 & 1804.36 \\
&  3 &   \nicefrac{7}{2} & 3d\nicefrac{5}{2} & 1783.72 &  0.05 &  0.02 & 1783.79 \\
&  0 &   \nicefrac{5}{2} & 3d\nicefrac{5}{2} & 1772.11 &  0.11 & -0.04 & 1772.18 \\[7pt]
$^{235}\text{U}$&  3 &   \nicefrac{7}{2} & 1s\nicefrac{1}{2} & 12175.51 &  6.83 &   0.00 & 12182.34 \\
&  7 &  \nicefrac{15}{2} & 1s\nicefrac{1}{2} & 11925.50&  4.66 &  0.00  & 11930.16 \\
&  9 &  \nicefrac{17}{2} & 1s\nicefrac{1}{2} & 11835.54&  3.54 &  0.00  & 11839.08 \\
&  3 &   \nicefrac{7}{2} & 2p\nicefrac{1}{2} & 6019.06 &  5.99 &  0.85 & 6025.90 \\
&  4 &   \nicefrac{7}{2} & 2p\nicefrac{1}{2} & 6015.01 &  5.96 &  0.83 & 6021.80 \\
&  4 &   \nicefrac{9}{2} & 2p\nicefrac{1}{2} & 5979.31 &  6.02 &  0.86 & 5986.19 \\
&  5 &   \nicefrac{9}{2} & 2p\nicefrac{3}{2} & 5928.94 &  6.06 &  0.88 & 5935.88 \\
&  6 &  \nicefrac{11}{2} & 2p\nicefrac{3}{2} & 5868.85 &  6.00 &  0.89 & 5875.74 \\
&  7 &  \nicefrac{15}{2} & 2p\nicefrac{1}{2} & 5798.66 &  5.30 &  0.91 & 5804.87 \\
&  8 &  \nicefrac{15}{2} & 2p\nicefrac{1}{2} & 5745.59 &  4.71 &  0.87 & 5751.17 \\
&  5 &   \nicefrac{7}{2} & 2p\nicefrac{3}{2} & 5673.10 &  3.12 & -0.42 & 5675.80 \\
&  6 &   \nicefrac{9}{2} & 2p\nicefrac{3}{2} & 5621.02 &  3.02 & -0.46 & 5623.58 \\
&  2 &   \nicefrac{7}{2} & 2p\nicefrac{3}{2} & 5620.12 &  2.78 & -0.56 & 5622.34 \\
&  9 &  \nicefrac{17}{2} & 2p\nicefrac{1}{2} & 5613.24 &  2.05 &  0.13 & 5615.42 \\
&  3 &   \nicefrac{9}{2} & 2p\nicefrac{3}{2} & 5586.28 &  2.81 & -0.54 & 5588.55 \\
&  7 &  \nicefrac{13}{2} & 2p\nicefrac{1}{2} & 5556.38 &  2.60 & -0.50 & 5558.48 \\
&  9 &  \nicefrac{15}{2} & 2p\nicefrac{3}{2} & 5493.59 &  2.44 &  0.24 & 5496.27 \\
&  8 &  \nicefrac{15}{2} & 2p\nicefrac{1}{2} & 5479.30 &  2.14 & -0.53 & 5480.91 \\
& 10 &  \nicefrac{17}{2} & 2p\nicefrac{3}{2} & 5393.16 &  1.77 &  0.13 & 5395.06 \\
&  9 &  \nicefrac{17}{2} & 2p\nicefrac{3}{2} & 5315.81 &  1.73 & -0.44 & 5317.10 \\
&  3 &   \nicefrac{7}{2} & 3d\nicefrac{3}{2} & 2767.16 &  0.44 &  0.09 & 2767.69 \\
&  1 &   \nicefrac{7}{2} & 3d\nicefrac{5}{2} & 2663.35 &  0.61 & -0.13 & 2663.83 \\
\end{tabular}
\end{ruledtabular}
\end{table*}
The quadrupole matrix elements from Eq.~\eqref{eq:quadel} can be calculated both for the rediagonalization in the dynamic hyperfine structure and for the evaluation of the residual second-order terms \eqref{eq:second}, using Eq.~\eqref{eq:matel}. As the next step, the total Hamiltonian~\eqref{eq:htot} is diagonalized in finite subspaces or modelspaces consisting of the muonic ($2p\nicefrac{1}{2}$, $2p\nicefrac{3}{2}$) or ($3d\nicefrac{3}{2}$, $3d\nicefrac{5}{2}$) doublet states and nuclear ground state rotational band. For Rhenium, the first six states with $I_N \in \{5/2,...,15/2\}$ are considered, and for Uranium with $I_N \in \{7/2,...,17/2\}$. The excitation energies of the nuclear states are summarized, along with other nuclear parameters, in Table~\ref{tab:params}. Thereby, the composite states and corresponding energies $E_{\text{quad}}$ from Eq.~\eqref{eq:compstate} are obtained and finally, for each of these states the residual second order quadrupole correction \eqref{eq:second} is calculated. Here, the intermediate sum goes over all nuclear and muonic states not included in the modelspace.
Note that for the muonic ground state, a rediagonalization is not necessary, since the diagonal matrix elements of the quadrupole interactions vanish for muonic states with $j=1/2$.
The quadrupole-Uehling contribution to the binding energies can be obtained by performing the calculations for the dynamic hyperfine structure twice; once with the matrix elements~\eqref{eq:quadel} and~\eqref{eq:quadel_uehl} containing both the electric and Uehling interaction ${V_{\text{el}}^{(2)}}{+}{V_{\text{uehl}}^{(2)}}$ and a second time only with the electric part ${V_{\text{el}}^{(2)}}$ from Eq.\eqref{eq:quadel}. The difference between those two approaches gives the quadrupole-Uehling corrections. Results for the residual second order quadrupole correction from Eq.~\eqref{eq:second} and for the quadrupole-Uehling corrections can be found in Table~\ref{tab:hfs} for a number of states.
\section{Discussion and Summary}
The quadrupole interaction in the framework of the dynamic hyperfine structure in heavy muonic atoms was analyzed by an fully relativistic treatment of the quadrupole-Uehling potential and of the residual second order terms.
The quadrupole-Uehling interaction was obtained by a multipole expansion of the Uehling potential for an arbitrary nuclear charge distribution without the assumption on the distance between muon and nuclear charge being large.
Since it has the same angular structure as the conventional quadrupole interaction, the quadrupole-Uehling expectation value vanishes between two muonic states with $j=1/2$, thus it does not affect the muonic ground state.
The calculations for Uranium show that it can lead to energy corrections almost on the keV level for very heavy nuclei with muonic $2p$ states and thus can be potentially relevant for comparison between theoretical predictions and experiments. Being a short-ranged potential, it falls of quickly for states further away from the nucleus. For states with $n\geq3$, we find values below $0.15\,$keV, even for high Z.
The generalization to Uehling corrections for higher order multipole is straight forward. In the case of muonic atoms, since the influence of higher order multipoles is already small, we expect this correction will not contribute significantly.

The residual second order quadrupole corrections in the dynamic hyperfine structure were calculated numerically using a basis of relativistic wave functions including nuclear finite size correction and Uehling correction due to the monopole part of the nuclear potential.
In contrast to the first order terms, the muonic ground state energy is affected by the second order corrections. Here, the energy correction amounts up to several keV. Also for muonic $2p$ states, it is of similar size. For the $3d$ levels, we find the energy corrections below half a keV, both for Rhenium and Uranium.
If a more complete nuclear model instead the rotational model is used, the additional nuclear states appear as intermediate state in the second order corrections, leading to the nuclear polarization corrections. Therefore, the approach presented in this work provides a basis for an accurate treatment of the muonic spectrum for the nuclear polarization effect in deformed muonic atoms.

\appendix
\section{Multipole expansion of electric \\and Uehling potential}
\label{sec:multipole}
With the multipole expansion of the Coulomb potential~\cite{jackson1999}
\begin{equation}
\frac{1}{|\vec{r}_\mu^{\,\prime}-\vec{r}_N^{\,\prime}|}=\sum_{l=0}^\infty \frac{r_<^l}{r_>^{l+1}}\sum_{m=-l}^l  \text{C}^*_{lm}(\vartheta_N^{\prime},\varphi_N^{\prime})\text{C}_{lm}(\vartheta_\mu^\prime,\varphi_\mu^\prime),
\end{equation}
where $r_<:=\min (r_\mu,r_N)$, $r_>:=\max (r_\mu,r_N)$, the electric potential~\eqref{eq:pot} can be written as
\begin{align}
V_{\text{el}}(\vec{r}_\mu^{\,\prime})=&\sum_{l,m}-Z\alpha \left[ \int \text{d}^3r^{\prime}_N \frac{r_<^l}{r_>^{l+1}}\text{C}^*_{lm}(\vartheta_N^{\prime},\varphi_N^{\prime})\rho(\vec{r}_N^{\,\prime})\right]\notag\\
&\,\times\text{C}_{lm}(\vartheta_\mu^{\prime},\varphi_\mu^\prime),
\end{align}
where ${\text{C}_{lm}(\vartheta,\varphi)}{=}{\sqrt{4\pi/(2l+1)}\text{Y}_{lm}(\vartheta,\varphi)}$ are the normalized spherical harmonics and primed coordinates belong to the nuclear body-fixed system.
For axially symmetric charge distributions, only the ${m}{=}{0}$-terms are non-zero after integrating over the charge distribution. The dependency on the muonic angular variables can be transformed to the laboratory system by
\begin{equation}
\text{P}_{l}(\cos\vartheta_\mu^\prime)=
 \sum_{m=-l}^l C^*_{lm}(\theta,\phi)C_{lm}(\vartheta_\mu,\varphi_\mu).
\end{equation}
Thereby, the potential as a function of the Euler angles and the muon's position in the laboratory frame reads
\begin{align}
V_{\text{el}}(\vec{r}_\mu,\phi,\theta)=&\sum_{l=0}^\infty-Z\alpha \left[ \int \text{d}^3r^{\prime}_N \frac{r_<^l}{r_>^{l+1}}\text{P}_{l}(\cos \vartheta_N^{\prime})\rho(\vec{r}_N^{\,\prime})\right]\notag\\
&\quad\,\times \sum_{m=-l}^l \text{C}^*_{lm}(\theta,\phi)\text{C}_{lm}(\vartheta_\mu,\varphi_\mu).\notag\\
=:&\sum_{l=0}^\infty Q_{\text{el}}^{(l)}(r_\mu)\sum_{m=-l}^lC^*_{lm}(\theta,\phi)C_{lm}(\vartheta_\mu,\varphi_\mu)&\notag\\
=:& \sum_{l=0}^\infty V^{(l)}_{\text{el}}(\vec{r}_\mu,\phi,\theta),
\label{eq:defmulti}
\end{align}
where $Q_{\text{el}}^{(l)}(r_\mu)$ describe the radial distribution of the mulitpole moments.

The Uehling potential can be expanded in multipoles in a similar way, now with a different dependence on $|\vec{r}_\mu^{\,\prime}-\vec{r}_N^{\,\prime}|$, as
\begin{align}
\frac{K_1(2m_e{|\vec{r}_\mu^{\,\prime} - \vec{r}^{\,\prime}\,|})}{|\vec{r}_\mu^{\,\prime} - \vec{r}^{\,\prime}\,|}
&=\sum_{l=0}^\infty c_l(r_\mu,r_N)\\
&\times\sum_{m=-l}^l \text{C}^*_{lm}(\vartheta_N^{\prime},\varphi_N^{\prime})\text{C}_{lm}(\vartheta_\mu^\prime,\varphi_\mu^\prime).\notag
\end{align}
The expansion coefficients $c_l$ can be calculated, using the fact that rotations do not change the absolute value of vectors, as
\begin{equation}
|\vec{r}_\mu^{\,\prime} - \vec{r}_N^{\,\prime}|
=|\vec{r}_\mu - \vec{r}_N|
=\sqrt{r_\mu^2 + r_N^2 - 2 r_\mu r_N y},
\end{equation}
with ${y}{=}{\cos(\sphericalangle \vec{r}_\mu\vec{r}_N)}$ being the cosine of the angle between the vectors $\vec{r}_\mu$ and $\vec{r}_N$, and read as
\begin{equation}
c_l(r_\mu,r_N)=\frac{2l+1}{2} \int_{-1}^1 \text{d}y \frac{K_1(2m_e{|\vec{r}_\mu - \vec{r}_N|})}{|\vec{r}_\mu - \vec{r}_N|} P_l(y),
\label{eq:defcl}
\end{equation}
and are evaluated by numerical integration.
Thereby, the Uehling potential can be written, analogously to Eq.~\eqref{eq:defmulti}, as
\begin{align}
V_{\text{uehl}}(\vec{r}_\mu,\phi,\theta)=&\sum_{l=0}^\infty-Z\alpha\frac{2\alpha}{3\pi} \notag\\
&\quad\times\left[ \int \text{d}^3r^{\prime}_N c_l(r_\mu,r_N)\text{P}_{l}(\cos \vartheta_N^{\prime})\rho(\vec{r}_N^{\,\prime})\right]\notag\\
&\quad\,\times \sum_{m=-l}^l \text{C}^*_{lm}(\theta,\phi)\text{C}_{lm}(\vartheta_\mu,\varphi_\mu).\notag\\
=:&\sum_{l=0}^\infty Q^{(l)}_{\text{uehl}}(r_\mu)\times\sum_{m=-l}^lC^*_{lm}(\theta,\phi)C_{lm}(\vartheta_\mu,\varphi_\mu)&\notag\\
=:& \sum_{l=0}^\infty V^{(l)}_{\text{uehl}}(\vec{r}_\mu,\phi,\theta).
\label{eq:defmultiuehl}
\end{align}
For $l=0$, the expression for Uehling potential of a spherical charge distribution~\cite{Fullerton1976} which only depends on $r_\mu$ is recovered as
\begin{align}
\label{eq:sph_uehl}
V^{(0)}_{\text{uehl}}(r_\mu)=& -\frac{2\alpha (Z\alpha)}{3 m_e r}\int_0^\infty \text{d}r^{\prime}\,\rho_0(r^\prime)\\
&\times\left[K_0(2m_e|r-r^\prime|)-K_0(2m_e(r+r^\prime))\right]\notag
\end{align}
where the spherically averaged part of the charge distribution is
\begin{equation}
\rho_0(r)=\int_0^{2\pi}\text{d}\varphi\int_0^\pi\text{d}\vartheta \sin(\vartheta)\rho(\vec{r}\,)/(4\pi).
\end{equation}
\section{Matrix elements\\of quadrupole interactions}
\label{sec:mat}
Both the electric and Uehling multipole interaction are a scalar product of rank-$l$ spherical tensors in the angular variables of the muon and of the nucleus. Thus, their matrix elements in states of defined total angular momentum can be reduced to a product of two reduced matrix elements~\cite{varshalovich1988} as
\begin{align}
&{\left< F_1M_1n_1\kappa_1I_1K\right|V^{(l)}(\vec{r}_\mu,\phi,\theta)\left|FMn_2\kappa_2I_2K\right>} {=}
{\delta_{\scriptscriptstyle F_1F_2}\delta_{\scriptscriptstyle M_1M_2}}\notag\\
&\qquad\times(-1)^{F_1+j_2+I_1}
\left\{
\begin{matrix}
j_1&j_2&l\\
I_1&I_2&F_1
\end{matrix}
\right\}\left< I_1K\big{|}\big{|}\text{C}_l(\theta,\phi)\big{|}\big{|}I_2K\right>\notag\\
&\qquad\times \left< n_1\kappa_1\big{|}\big{|}Q^{(l)}(r_\mu)\text{C}_l(\vartheta_\mu,\varphi_\mu) \big{|}\big{|}n_2\kappa_2\right>.\label{eq:matel}
\end{align}
Here, $\kappa_i$ is related to the total angular momentum as $j_i=|\kappa_i|-1/2$.
The matrix elements for the nucleus, reduced in $M$ but not in $K$, read~\cite{brown_carrington}
\begin{align}
\left< I_1K\big{|}\big{|}\text{C}_l(\theta,\phi)\big{|}\big{|}I_2K\right>=&(-1)^{I_2+K}
\sqrt{(2I_1+1)(2I_2+1)}\notag\\
&\times\,\left(
\begin{matrix}
I_1&I_2&l\\
-K&K&0
\end{matrix}
\right),
\end{align}
and the reduced matrix elements in the muonic variables are~\cite{johnson2007}
\begin{align}
&\left< n_1\kappa_1\big{|}\big{|}Q^{(l)}(r_\mu)\text{C}_l(\vartheta_\mu,\varphi_\mu) \big{|}\big{|}n_2\kappa_2\right> =  (-1)^{j_1+1/2}\\
&\quad \times \sqrt{(2j_1+1)(2j_2+1)} \pi(l_1+l_2+l)
\left(
\begin{matrix}
j_1&j_2&l\\
-\frac{1}{2}&\frac{1}{2}&0
\end{matrix}
\right)\notag\\
&\quad\times\,
\int \text{d}rr^2 ({g_{n_1\kappa_1}(r)g_{n_2\kappa_2}(r)}{+}{f_{n_1\kappa_1}(r)f_{n_2\kappa_2}(r)}) Q^{(l)}(r_\mu),\notag
\end{align}
where $g_{n\kappa}(r)$ and $f_{n\kappa}(r)$ are the two radial functions of the solutions of the Dirac equation in the spherically symmetric potential~\cite{greiner2000} from Eq.~\eqref{eq:h0mu}. The orbital angular momentum quantum number is  $l_i=|\kappa_i|+(\mathrm{sgn}(\kappa_i)-1)/2$, and the parity selection rule is included by the function
\begin{equation}
\label{eq:parityFunc}
\pi(x) =
\left\{
\begin{matrix}
1\,\text{, x even;}\phantom{11}\\
0\,\text{, otherwise.}
\end{matrix}
\right.
\end{equation}
%
%
%
\end{document}